\documentclass[twocolumn,showpacs,preprintnumbers,amsmath,amssymb,prl,superscriptaddress]{revtex4}

\usepackage{graphicx}
\usepackage{color}

\usepackage{hyperref}

\begin{document}

\title{Revealing the dual nature of magnetism in iron pnictides and iron chalcogenides using x-ray emission spectroscopy}

\author{H.~Gretarsson}
\author{A.~Lupascu}
\affiliation{Department of Physics, University of Toronto, 60
St.~George St., Toronto, Ontario, M5S 1A7, Canada}
\author{Jungho~Kim}
\author{D.~Casa}
\author{T.~Gog}
\affiliation{Advanced Photon Source, Argonne National Laboratory,
Argonne, Illinois 60439, USA}
\author{W.~Wu}
\author{S.~R.~Julian}
\affiliation{Department of Physics, University of Toronto, 60
St.~George St., Toronto, Ontario, M5S 1A7, Canada}
\author{Z.~J.~Xu}
\author{J.~S.~Wen}
\author{G.~D.~Gu} \affiliation{CMP\&MS Department, Brookhaven
National Laboratory, Upton, New York 11973, USA}
\author{R.~H.~Yuan}
\author{Z.~G.~Chen}
\author{N.-L.~Wang}
\affiliation{Beijing National Laboratory for Condensed Matter Physics, Institute of Physics, Chinese Academy of Sciences, Beijing 100190, China}
\author{S.~Khim}
\affiliation{CeNSCMR, Department of Physics and Astronomy, Seoul National University, Seoul 151-747, Republic of Korea}
\author{K.~H.~Kim}
\affiliation{CeNSCMR, Department of Physics and Astronomy, Seoul
National University, Seoul 151-747, Republic of Korea}
\author{M.~Ishikado}
\author{I.~Jarrige}
\author{S. Shamoto}
\affiliation{Quantum Beam Science Directorate, Japan Atomic Energy
Agency, Tokai, Naka, Ibaraki 319-1195, Japan}
\author{J.-H. Chu}
\author{I. R. Fisher}
\affiliation{Geballe Laboratory for Advanced Materials and
Department of Applied Physics, Stanford University, Stanford,
California 94305, USA}
\author{Young-June~Kim}
\email{yjkim@physics.utoronto.ca} \affiliation{Department of
Physics, University of Toronto, 60 St.~George St., Toronto, Ontario,
M5S 1A7, Canada}

\date{\today}

\begin{abstract}
We report Fe K$\beta$ x-ray emission spectroscopy study of local
magnetic moments in various iron based superconductors in their
paramagnetic phases. Local magnetic moments are found in all samples
studied: PrFeAsO, $\rm Ba(Fe,Co)_2As_2$, LiFeAs, Fe$_{1+x}$(Te,Se),
and $\rm A_2Fe_4Se_5$ (A=K, Rb, and Cs). The moment size varies
significantly across different families. Specifically, all iron
pnictides samples have local moments of about 1~$\mu_B$/Fe, while
FeTe and $\rm K_2Fe_4Se_5$ families have much larger local moments
of $\sim 2 \mu_B$/Fe, $\sim 3.3 \mu_B$/Fe, respectively. In
addition, we find that neither carrier doping nor temperature change
affects the local moment size.
\end{abstract}



\maketitle

The duality of local moment -- itinerant electron in magnetism has
long been confounding researchers trying to explain metallic
ferromagnets such as Fe and Ni
\cite{Kakizaki1994,Lichtenstein2001,Pickel2010}, and it is again at
the center of debate regarding microscopic understanding of
magnetism in the iron based superconductors
\cite{Johannes2009,Lumsden2010,Johnston2010,Hu2011,Yin2011}. Various
theoretical studies have approached magnetism in these materials
from the itinerant viewpoint. In particular, density functional
theory (DFT) predicted spin-density wave type magnetic order in
La(O$_{1-x}$F$_x$)FeAs \cite{Singh2008,Dong2008,Mazin2008}, which
was later confirmed by neutron scattering experiments
\cite{Cruz2008}. Despite this success, fully itinerant description
seems to be insufficient. For example, DFT calculation consistently
overestimates the ordered magnetic moment of the parent compounds of
iron pnictides. The theoretical value of $\sim 2 \mu_B$ (per Fe atom
throughout this Letter) \cite{Cao2008} is much larger than the
ordered moment determined from neutron diffraction experiments
\cite{Lumsden2010}. Such a discrepancy has been attributed to the
magnetic frustration and fluctuation effects from the local moment
perspective \cite{Yildirim2008,Si2008}. Perhaps more pertinent to
our discussion of the dual nature is the magnetic behavior in the
paramagnetic regime. In the purely itinerant picture, local magnetic
moment would disappear above the transition temperature in zero
field (Pauli paramagnet), while in the local picture (Curie
paramagnet), local moments would be fluctuating and pointing in
random directions. Therefore, the presence of local moments in the
paramagnetic phase would be a telltale sign of localized magnetism.

However, experimentally probing local magnetic moments is
challenging. The temperature dependence of spin susceptibility
measured with magnetometry or NMR typically is strongly affected by
the spin correlation, especially in iron pnictides and
chalcogenides, in which magnetic interaction energy scale is quite
large \cite{Johnston2010}. Neutron scattering is useful, and has
been used to detect local moments \cite{Diallo2010,Igor2011,Xu2010},
but usually it is time consuming and requires large quantity of
sample. Here we introduce X-ray Emission Spectroscopy (XES), which
is a bulk-sensitive method to detect local magnetic moment of Fe
\cite{Rueff1999,Lin2007,Vanko2006,Bergmann2009,Rueff2008}. This XES
technique is widely used in earth sciences to probe spin states of
iron in minerals \cite{Lin2007}. A recent development of
quantitative analysis method has made it possible to obtain local
magnetic moment information without detailed lineshape analysis
\cite{Vanko2006}.

In this Letter, we report our comprehensive XES investigation of
magnetic moments in a number of iron pnictides and iron
chalcogenides: PrFeAsO, Ba(Fe,Co)$_2$As$_2$, LiFeAs,
Fe$_{1+x}$(Te,Se), and $\rm A_2Fe_4Se_5$ (A=K, Rb, and Cs)
\cite{composition-note}. We find that local moments are present at
room temperature in all samples studied. Furthermore, the size of
the local moments vary significantly among the samples studied,
ranging from 0.9 $\mu_B$ in LiFeAs to 3.3 $\mu_B$ in $\rm
K_2Fe_4Se_5$. This result suggests that the magnetism in iron based
superconductors requires a description taking into account the local
moment as well as the Fermi surface nesting. The relative importance
of local moment versus itinerant magnetism depends on the type of
anions and the structural details. Specifically, $\rm A_2Fe_4Se_5$
is almost entirely described by local moments, while the local
moment size decreases for Fe$_{1+x}$(Te,Se), and greatly suppressed
for the pnictides samples. However, the variation of the magnetic
moment size among the Fe pnictides (111, 122, and 1111) is found to
be very small. We also discuss the possible origin of such a
material dependence in view of the recent theoretical study by Yin
{\it et al.}, in which magnetic moments were discussed in relation
to orbital occupancy \cite{Yin2011}.

The x-ray emission spectroscopy (XES) was performed at the Advanced
Photon Source on the undulator beamline 9ID-B. The beam was
monochromatized by a double-bounce Si(111) crystal and a Si(311)
channel-cut secondary crystal. A spherical (1 m radius) diced
Ge(620) analyzer was used to obtain an overall energy resolution of
0.4 eV (FWHM of elastic line). The energy calibration was based on
the absorption spectrum through a thin Fe-foil, and incident x-ray
energy of 7.140 keV was used. Use of such a hard x-ray ensures that
the spectra are not surface sensitive. Details of the growths and
characterization of the single crystal samples have been reported in
earlier publications
\cite{Chu2009,Ishikado2009,BSL2010,ZChen2011,Wu2011}. All
measurements were carried out at room temperature, except for the
temperature dependence study, for which a closed-cycle cryostat was
used.

\begin{figure}
\includegraphics[width=\columnwidth]{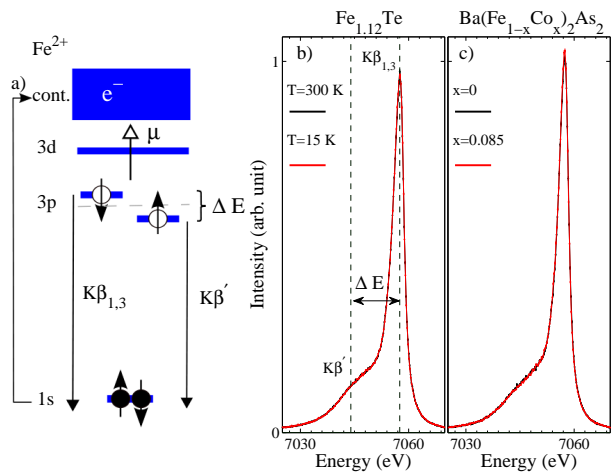}
\caption{(Color online) (a) Schematic diagram of the Fe K$\beta$
emission process in the atomic limit for Fe$^{2+}$. The 3p core-hole
in the final state interacts with the net magnetic moment
$\vec{\mu}$ in the 3d valence shell, creating two different final
states K$\beta_{1,3}$ and K$\beta^\prime$  with opposite core-hole
spins, separated in energy by $\Delta E$. (b) K$\beta$ emission line
for Fe$_{1.12}$Te taken above and below T$_N$ = 58. The splitting,
$\Delta E$, between K$\beta_{1,3}$ and K$\beta^\prime$ is caused by
the local magnetic moment.(c) K$\beta$ emission line for
BaFe$_2$As$_2$ for different Co doping. } \label{fig01}
\end{figure}


The local moment sensitivity of the K$\beta$ emission line (3p
$\rightarrow$ 1s) originates from a large overlap between the 3p and
3d orbitals. In Fig. \ref{fig01}(a) we show a schematic diagram of
the process for Fe$^{2+}$ in the atomic limit. The K$\beta$ emission
process has a core-hole in the final state (3$p^5$) which interacts
strongly with the 3$d^6$ valence electrons, affecting the possible
final state configurations of the K$\beta$ spectra
\cite{Tsutsumi1976,Peng1994}. In particular, such exchange
interactions are mainly driven by the presence of a net magnetic
moment in the 3d valence shell, resulting in final states with
antiparallel or parallel net spins between the 3$p^5$ core-hole and
3$d^6$ valence shell, as shown in Fig. \ref{fig01}(a). Since the
3p-3d interaction is local, this method is not sensitive to the
long-range order, but only probes local magnetic moment. The two
main multiplet features can be recognized in the K$\beta$ emission
line as the main peak K$\beta_{1,3}$ and the low energy satellite
K$\beta^\prime$, respectively. An example of such a splitting in the
K$\beta$ emission line for Fe$_{1.12}$Te is seen in Fig.
\ref{fig01}(b), in which the splitting between the two features,
$\Delta E$, was found to be $\sim$ 13.25 eV. The size of this
splitting depends on the local moment \cite{Tsutsumi1976}, but
actually extracting the satellite peak position from fitting is
quite difficult for a system with weak moment [see Fig.~1(c)]. In
their study of the 3s core level emission from $\rm
CeFeAsO_{0.89}F_{0.11}$, Bondino {\it et al.} used this method to
obtain about $1~\mu_B$ for the local moment size in this sample
\cite{Bondino2008}.

\begin{figure*}[htb]
\includegraphics[width=1.8\columnwidth]{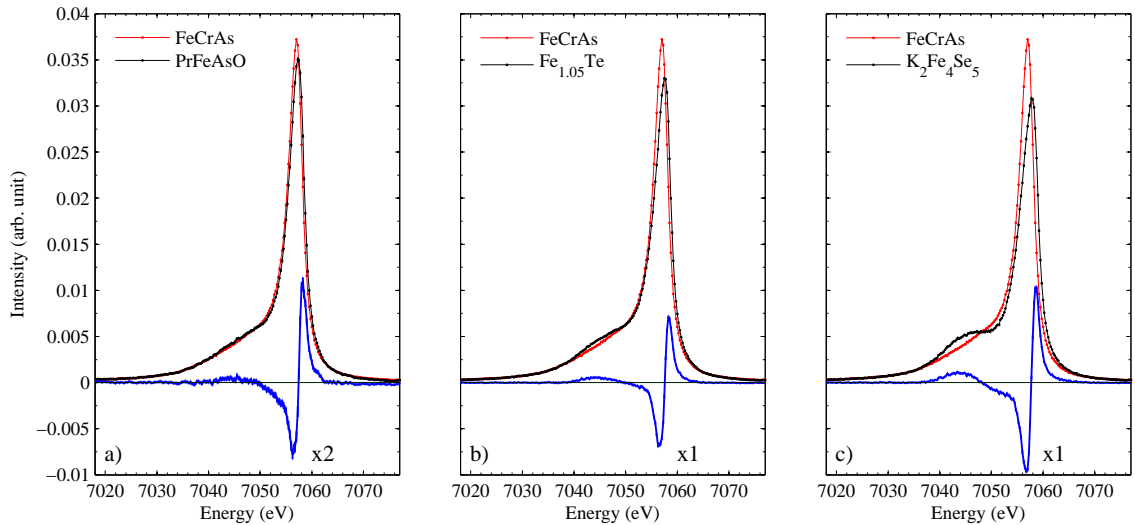}
\caption{(Color online) The XES spectra of the Fe K$\beta$ emission
lines for (a) PrFeAsO, (b) Fe$_{1.05}$Te, and (c) $\rm K_2Fe_4Se_5$.
The nonmagnetic reference spectra of FeCrAs, and the difference
spectra are also plotted. Note that the difference spectrum for
PrFeAsO was magnified by a factor of two.} \label{fig02}
\end{figure*}

Recently, a quantitative method based on the integration of spectral
weight difference has been suggested as a way to determine the local
moment \cite{Vanko2006}. Since both the intensity of the satellite
and the splitting $\Delta E$ are related to the 3d local moment
\cite{Tsutsumi1976}, this integrative method utilizes the whole
spectrum and not just the peak position. The method has been
successfully used in a number of applications
\cite{Lin2007,Bergmann2009,Rueff2008}. In order to quantitatively
derive the total local moment from the K$\beta$ line using the
integrated absolute difference (IAD) analysis, one needs to have a
reference sample with the same local coordination around Fe, but
with Fe ion in the non-magnetic low-spin (LS) state. The IAD is then
the integrated absolute difference between the spectrum measured and
the non-magnetic reference spectra. Vanko {\em et~al.}
\cite{Vanko2006} showed that the IAD is linearly proportional to the
spin magentic moment of the Fe atom. For this purpose, we use FeCrAs
as a non-magnetic reference sample. The Fe atoms in FeCrAs is
tetrahedrally coordinated with As, as is found in Fe
superconductors. Although FeCrAs orders magnetically, both
experimental \cite{Wu2009} and theoretical \cite{Ishida1996} studies
have shown that the magnetism entirely resides on the Cr sites, and
Fe is non-magnetic. In order to determine the absolute scale of the
magnetic moment, we use the value for Fe-chalcogenide $\rm
K_2Fe_4Se_5$. Since this is an insulating sample, we assume that the
local moment size is the same as the ordered magnetic moment at room
temperature; both neutron scattering \cite{WeiBao2011} and DFT
calculation \cite{Yan2011} results agree on the value of ordered
magnetic moment of 3.3 $\mu_B$.

In Fig. \ref{fig02} we show representative K$\beta$ XES data for (a)
PrFeAsO, (b) Fe$_{1.05}$Te, and (c) $\rm K_2Fe_4Se_5$ along with the
FeCrAs spectrum. To follow the procedure from Ref.~\cite{Vanko2006},
the area underneath each spectrum was normalized to unity.  The
reference spectrum is then subtracted from the sample spectrum, and
the resulting difference is plotted. The IAD quantity is extracted
by integrating the absolute value of this difference spectrum. What
is evident from Fig. \ref{fig02} is that the intensity of
K$\beta^\prime$ changes quite a bit going from PrFeAsO to $\rm
K_2Fe_4Se_5$. In addition we see a shift of the main K$\beta_{1,3}$
peak position towards higher energy as K$\beta^\prime$ increases in
intensity, a further evidence of the local moment variation
\cite{Vanko2006}.

\begin{figure}[htb]
\includegraphics[width=\columnwidth]{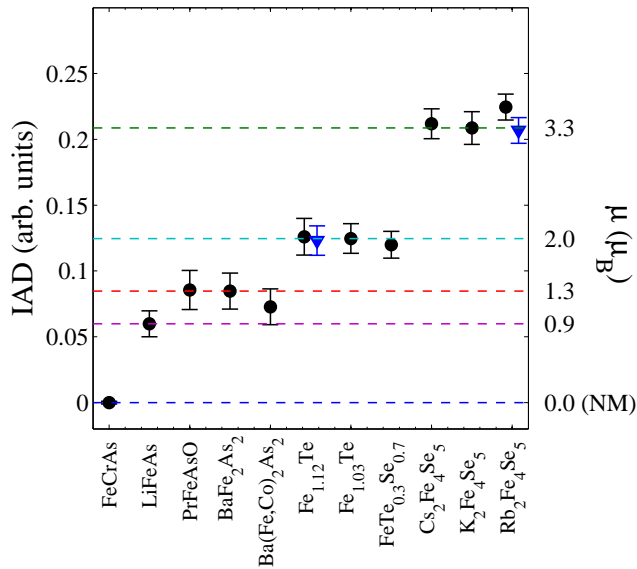}
\caption{(Color online) The IAD values derived from the XES spectra
for various samples. The room temperature data are shown in circles,
and the low-temperature IAD values at T = 15 K are shown in
triangles for Fe$_{1.12}$Te and $\rm Rb_2Fe_4Se_5$. On the right
hand side is the local magnetic moment (${\rm \mu}$)
scale.}\label{fig03}
\end{figure}

The IAD values so-obtained for all the samples are plotted in Fig.
\ref{fig03}. On the right hand side of the figure is the local
moment scale determined from the $\rm K_2Fe_4Se_5$ ordered moment
\cite{WeiBao2011}. The moment sizes roughly falls within three
groups. All $A \rm Fe_2Se_2$ samples have approximately the same
moment size close to 3.3 $\mu_B$, while the local moment size for
all Fe(Te,Se) samples is around 2 $\mu_B$. Both of these values are
close to the respective ordered moment size, but much larger than
the values for Fe pnictides, which carry local moments of about 1
$\mu_B$. This latter value is quite similar to the ordered moment
reported for BaFe$_2$As$_2$ ( 0.9 ${\rm \mu_B}$ \cite{Wilson2009}),
but much larger than the values for PrFeAsO and LiFeAs. The ordered
moment size for PrFeAsO is 0.35 $\mu_B$ and LiFeAs does not order
magnetically; isostructural NaFeAs has ordered moment of 0.09
$\mu_B$.

We also studied the temperature and carrier doping dependence of the
local moment size or lack thereof. In Fig.~\ref{fig01}(b), we show
XES spectra for Fe$_{1.12}$Te obtained at two different temperatures
above and below the magnetic ordering transition ($T_N \approx
58$~K). In Fig.~\ref{fig01}(c),
Ba(Fe$_{0.915}$Co$_{0.085}$)$_2$As$_2$ XES spectrum is compared with
that of undoped $\rm BaFe_2As_2$ compound. Magnetic order is
suppressed in the  Ba(Fe$_{0.915}$Co$_{0.085}$)$_2$As$_2$ sample,
which is superconducting with $T_c \approx$ 17 K. The lack of any
change in both figures indicates that the local moment size is
insensitive to the presence of long-range order or carrier
concentration. Similar conclusion can be reached from additional
temperature and doping dependence studies for $\rm Rb_2Fe_4Se_5$ and
$\rm FeTe_{0.3}Se_{0.7}$ (included in Fig.~\ref{fig03}). This is in
contrast to recent neutron scattering results, in which increased
moment size in the paramagnetic phase of Fe$_{1.1}$Te was observed
\cite{Igor2011}. In the case of $\rm Rb_2Fe_4Se_5$ the lack of
change above and below the superconducting T$_c$ = 30 K, suggests
that a large local moment ($\sim$ 3.3 ${\rm \mu_B}$) does exist in
the superconducting phase, although this could be due to a phase
separation as suggested in a recent study \cite{Chen2011}.

Summarizing our experimental findings, local magnetic moments are
found in the paramagnetic phase of all Fe pnictides and
chalcogenides samples. In addition, the local moment size only
depends on which anion the sample has, and is insensitive to doping
and temperature. In particular, we find that the local moment size
varies very little among the three ferropnictides families, despite
widely different ordered moment size. In their recent dynamical mean
field theory (DMFT) calculation combined with DFT, Yin et al. found
that the paramagnetic fluctuating local moment was rather sample
independent among the `111', `1111', and `122' families of iron
pnictides \cite{Yin2011}, which is consistent with our experimental
observation. However, the fluctuating local moment from the
calculation ($\sim$ 2.4 ${\rm \mu_B}$) is still larger than the
observed 1 $\mu_B$ for ferropnictides, implying that there exist
``missing" magnetic moments. We speculate that XES is weighted such
that local electrons are emphasized while more itinerant electrons
are not properly counted, due to the local nature of the core-hole
potential. Yin and coworkers indeed found that $t_{2g}$ electrons
have more local character in ferropnictides \cite{Yin2011}.

On the other hand, the calculated fluctuating moment size of the
`11' iron chalcogenides agrees fairly well with our XES value ($\sim
2 \mu_B$). In addition, the local moment size was found to be
similar in both FeTe and FeSe, even though the long range order is
lost in FeSe. These results are in agreement with our results in
Fig. \ref{fig03} in which no difference was seen in the IAD
valuesfor Fe$_{1.12}$Te, Fe$_{1.05}$Te and FeTe$_{0.3}$Se$_{0.7}$.
Yin and coworkers suggested that the structural details of the
Fe-pnictogen/chalcogen tetrahedra are crucial in determining the
orbital occupancy and the quasiparticle mass enhancement, which in
turn determines the magnetic moment \cite{Yin2011}. In particular,
the large Te ions make this system structurally distinct from
ferropnictides.

In conclusion, we find that PrFeAsO, Ba(Fe,Co)$_2$As$_2$, LiFeAs,
Fe$_{1+x}$(Te,Se), and $\rm A_2Fe_4Se_5$ (A=Cs, K, and Rb) all
possess local magnetic moments even in their paramagnetic phases. By
analyzing our x-ray emission spectroscopy data using recently
developed integrated absolute difference method, we could determine
the local moment size for each sample. The local moment size of iron
chalcogenides agree with theoretical calculation values and
experimentally measured static moment size. However, the local
moment size of ferropnictides is universally around 1 $\mu_B$, which
could originate from the more localized $t_{2g}$ electrons. Our
results perhaps suggest that it is the $t_{2g}$ local moment that
orders in ferropnictides, eliminating the need for Fermi-surface
nesting, as argued in a recent theoretical study
\cite{Johannes2009}.

\acknowledgements{We would like to thank H. Eisaki and A. Iyo for
fruitful discussion and technical assistance of the PFAO crystal
growth. Research at the U. of Toronto was supported by the NSERC,
CFI, OMRI, and CIfAR. Y.-J.K. was supported by the KOFST through the
Brainpool program. Use of the APS was supported by the U. S. DOE,
Office of Science, Office of BES, under Contract No.
W-31-109-ENG-38. The work at BNL was supported by DOE under contract
No. DE-AC02-98CH10886. N.-L. W. acknowledges NSFC and MOST 973
project from China. Work at SNU was supported by National Creative
Research Initiative(2010-0018300). Work at JAEA was supported by
JST, TRIP. Work at Stanford was supported by the DOE-BES under
contract DE-AC02-76SF00515.}

\end{document}